\documentclass[useAMS,usenatbib,usegraphicx]{mn2e}

\title[Optical and near-infrared colour distributions of early-type
galaxies in ABELL 2199]{Radial Variation of Optical and Near-Infrared
Colours in Luminous Early-Type Galaxies in ABELL 2199}

\author[N. Tamura and K. Ohta]{Naoyuki
Tamura$^{1,3}$\thanks{E-mail:naoyuki.tamura@durham.ac.uk} and Kouji
Ohta$^{2,3}$\\ $^{1}$Department of Physics, University of Durham, South
Road, Durham, DH1 3LE, UK\\ $^{2}$Department of Astronomy, Kyoto
University, Kyoto 606-8502, Japan\\ $^{3}$Visiting astronomer of UKIRT
and the University of Hawaii 2.2m telescope.}

\begin{document}

\date{}

\pagerange{\pageref{firstpage}--\pageref{lastpage}} \pubyear{2004}

\maketitle

\label{firstpage}

\begin{abstract}

We performed $K$ band surface photometry for luminous early-type
galaxies in a nearby rich cluster ABELL 2199. Combining it with $B$ and
$R$ band surface photometry, radial variations of $B-R$ and $R-K$
colours in the galaxies were investigated. It is found that the inner
regions of the galaxies are redder in both of $B-R$ and $R-K$ colours.
Comparing the radial variations of both of the colours with predictions
of Simple Stellar Population (SSP) models for a range of ages and
metallicities, it is suggested that the cluster ellipticals have
negative metallicity gradients but their age gradients are consistent
with zero, although our sample is small; the typical metallicity
gradient is estimated to be $ -0.16 \pm 0.09$ in $d \log Z /d \log r$,
while the age gradient is estimated to be $-0.10 \pm 0.14$ in $d \log
{\rm (age)} /d \log r$.
Considering that similar results have also been derived in the other
recent studies using samples of ellipticals in the Coma cluster and less
dense environments, it seems that there is no strong dependence on
galaxy environment in radial gradient of stellar population in
elliptical galaxy.

\end{abstract}


\section{INTRODUCTION}

Elliptical galaxies in a cluster show a tight colour-magnitude (CM)
relation; colours of more luminous ellipticals are systematically redder
(e.g., Bower, Lucey, \& Ellis 1992). It has been found that the relation
still holds in distant rich clusters around $z \sim 1$ (e.g., Stanford,
Eisenhardt, \& Dickinson 1998). Based on these observational results,
galaxy evolution models taking into account chemical evolution claim
that elliptical galaxies in rich clusters formed at, e.g., $z > 3$
through a monolithic collapse of a gas cloud (e.g., Kodama \& Arimoto
1997).
In this monolithic formation, a radial variation of stellar population
in the sense that stellar metallicity is higher on average towards the
galaxy centre ({\it metallicity gradient} hereafter) is expected to form
as a result of initial starburst and subsequent blow of galactic wind; a
more extended period of active star formation and thus more chemical
enrichment is expected in the inner portions of a galaxy (e.g., Kawata
1999).
Indeed, many elliptical galaxies are known to have radial gradients of
colours and absorption line strengths: colours are redder and metal
absorption lines are stronger in the inner regions (e.g., Peletier et
al. 1990; Davies, Sadler, \& Peletier 1993), and the evolution of the
colour gradients investigated by looking at distant ellipticals suggests
that these gradients originate from the metallicity gradients (Tamura et
al. 2000; Tamura \& Ohta 2000; Saglia et al. 2000; La Barbera et al.
2003).

On the other hand, a bottom-up structure formation based on the Cold
Dark Matter (CDM) universe generally succeeds in reproducing many
observational aspects of galaxies in the present day universe. In this
scheme, an elliptical galaxy is considered to form via a major merger of
galaxies. Accordingly, even if the progenitors of an elliptical galaxy
have some radial variations of stellar population, the variations are
more likely to be smeared out during the merger. It is also expected
that the histories of mass assembly and star formation of an elliptical
galaxy can be different from galaxy to galaxy, even if elliptical
galaxies having similar masses at a given redshift are considered.
Therefore, the radial gradient of stellar population in an elliptical
galaxy would not be expected to correlate with the galaxy mass, although
there is a possibility that the gradients in more massive ellipticals
are less steep because they are likely to experience more merging events
on average than less massive systems.

These considerations show that investigating colour or absorption line
strength gradients in elliptical galaxies in a statistical manner is a
potentially powerful probe of galaxy formation process (Kobayashi 2004).
We recently performed $B$ and $R$ bands surface photometry for
elliptical and S0 galaxies (E/S0s) in ABELL 2199, which is one of nearby
rich clusters, to study their colour gradients (Tamura \& Ohta 2003;
hereafter Paper I).
From these optical data, it is found that (1) the average metallicity
gradient in the ellipticals estimated from their colour gradients on the
assumption of no age gradient is $\sim -0.3 \pm 0.1$ in $d \log Z / d
\log r$, which can be reproduced by a recent model of the
monolithic-like galaxy formation (e.g., Kawata 1999), and (2) for the
galaxies brighter than $R = 15$ mag ($L \sim L^{*}$ at the distance of
ABELL 2199), more luminous ones tend to have steeper colour gradients
(Figure \ref{corr}). These results are consistent with the monolithic
collapse scenario rather than the bottom-up scheme (Larson 1974;
Carlberg 1984; Kawata \& Gibson 2003).  Considering that almost all the
other previous studies have targeted ellipticals in environments less
dense than ABELL 2199 and no such correlations have been found in the
previous studies, our data may imply an environmental dependence of
formation process of elliptical galaxy.

However, it is premature to conclude that the cluster E/S0s formed
through the monolithic collapse scenario, and it needs to be confirmed
that the galaxy luminosities correlate with the metallicity gradients.
Although this requires to disentangle the age--metallicity degeneracy
which is very hard in practice, one possible way to address this issue
is to add another colour information including a near-infrared (NIR)
band to the optical colour and see the radial variations of both of the
two colours (e.g., Peletier, Valentijn, \& Jameson 1990). It will allow
us to see whether the colour variation is consistent with a pure
metallicity gradient or a significant contribution of age gradient is
essential to explain it.
In this paper, we present results of $K$ band surface photometry for the
E/S0 galaxies in ABELL 2199 to be combined with the optical surface
photometry (Paper I). $K$ band is chosen to make a baseline of
wavelength longer to break the degeneracy more clearly. Among the
luminous ellipticals plotted in Figure \ref{corr}, we observed 8
galaxies. We briefly summarise optical observation and then describe NIR
observation and data reduction in the next section. Data analysis and
results are presented in \S~3. After discussing the results in \S~4, we
summarise this paper in \S~5.

\section{OBSERVATION AND DATA REDUCTION}

\subsection{Overall Description of Data Acquisition}

\begin{figure*}
 \centering
 \includegraphics[height=16cm,angle=-90,keepaspectratio,clip]{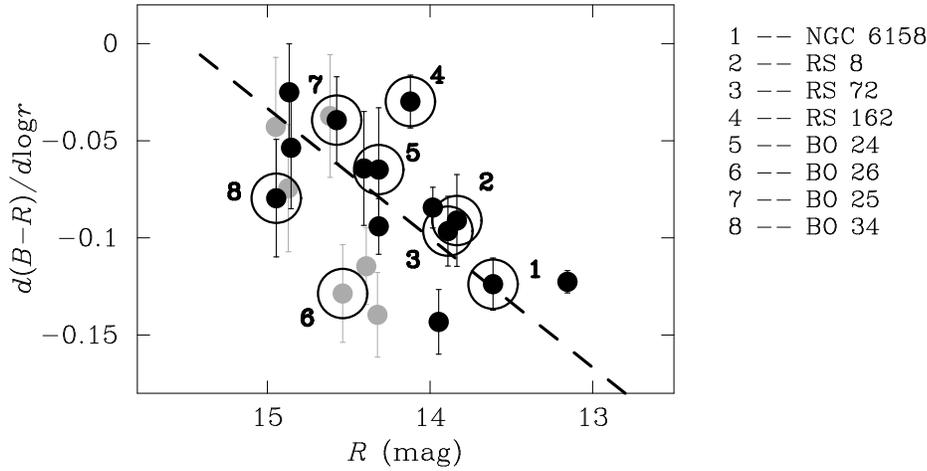}
 \caption{The relationship between $B-R$ colour gradients and $R$ band
 total magnitudes of luminous ($R \la 15$ mag or $L \ga L^{*}$)
 early-type galaxies in ABELL 2199 (Paper I). Black and grey symbols
 indicate galaxies with $r_e \ga 3^{\prime\prime}$ and those with 
 $r_e \la 3^{\prime\prime}$, respectively. Dashed line indicates the trend
 seen in this plot, which becomes clearer for the galaxies with
 $r_e \ga 3^{\prime\prime}$. The $K$ band data were newly added for the
 numbered galaxies with circles in this plot and these galaxies are
 discussed in this paper.} \label{corr}

\end{figure*}

We started studying colour gradients in the cluster ellipticals with
optical data, which is described in Paper I in detail. We performed $B$
and $R$ bands surface photometry for 40 early-type (E, E/S0, and S0)
galaxies in ABELL 2199 selected from the catalog by Lucey et al.
(1997). The galaxy morphology listed by Lucey et al. (1997) is mostly
taken from Butcher \& Oemler (1985) and Rood \& Sastry (1972).
Although studying these morphologically selected early-type galaxies
could be one option, it is known that there is a variety in
spheroidal-to-total luminosity ratio among galaxies in each
morphological type (Simien \& de Vaucouleurs 1986) and some of them can
be disk dominated galaxies. In order to securely sample galaxies
dominated by spheroidal components, we made azimuthally averaged radial
surface brightness profiles of the galaxies and performed decompositions
of the profiles into bulge and disk components to estimate
bulge-to-total luminosity ratios ($B/T$s). We sampled galaxies with
$B/T$s in $B$ band larger than 0.6. This criterion isolates galaxies
earlier than E/S0 or $T \lid -3$ according to Simien \& de Vaucouleurs
(1986), where $B/T$s of nearby galaxies are also derived in $B$ band.
Eventually, 31 galaxies were sampled out of the 40 galaxies observed.
The galaxies plotted in Figure \ref{corr} are those brighter than $R=15$
mag ($L \sim L^{\ast}$) and 8 of these galaxies were observed in $K$
band.

\subsection{The optical observation}

Since the details of the optical observation and data are described in
Paper I, we briefly summarise them here. The imaging observations in $B$
and $R$ bands were performed on 20 and 21 June 2001 with Tektronix
2K$\times$2K CCD on the University of Hawaii 2.2m telescope. One pixel
of the CCD subtends $0_{\cdot}^{\prime\prime}22$ on sky and thus each
frame covers a field of view of $\sim
7_{\cdot}^{\prime}5\times7_{\cdot}^{\prime}5$. A typical seeing size
during the observing run was $\sim 0_{\cdot}^{\prime\prime}9$ in FWHM. A
total exposure time at each field is typically 1,800 sec in $B$ band and
750 sec in $R$ band, each of which was divided into three or four
exposures with the telescope dithered.

The imaging data were reduced with IRAF\footnote{IRAF is distributed by
the National Optical Astronomy Observatories, which is operated by the
Association of Universities for Research in Astronomy, Inc. under
cooperative agreement with the National Science Foundation} in the
standard procedure. After bias subtraction and flat-fielding by
dome-flat frames, sky subtraction was carried out in each frame and
the sky-subtracted frames at each field were registered by sub-pixel
shifts. The FWHMs of the point spread functions (PSFs) were matched by
convolutions of images with Gaussian kernels and these images were
combined with the 3 $\sigma$ clipping algorithm. Finally a PSF size of
the stacked image in one band was matched to that in another band by
smoothing with Gaussian kernels. A resulting PSF size is
$1_{\cdot}^{\prime\prime}1$ in FWHM.

\setlength{\tabcolsep}{1mm}
\begin{table*}
 \centering
 \begin{minipage}{160mm}
 \begin{tabular}{lcccccccccc}
 \hline
  \multicolumn{1}{c}{Galaxy} & $R$   & $d(B-R)/d\log r$ &
  $d(R-K)/d\log r$  & $r_e$          & $r_{\rm in}/r_e$ &
  $r_{\rm out}/r_e$ & $B-R$          & $R-K$            &
  $\Delta (B-R)$    & $\Delta (R-K)$ \\
                    & (mag)          & (mag/dex)        &
  (mag/dex)         & (arcsec)       &                  &
                    & (mag)          & (mag)            &
  (mag)             & (mag)          \\
  \multicolumn{1}{c}{(1)}    & (2)   & (3)              &
  (4)               & (5)            & (6)              &
  (7)               & (8)            & (9)              &
  (10)              & (11)           \\
  \hline
NGC 6158 & 13.6 & $-0.12 \pm 0.01$ & $-0.17 \pm 0.01$ & 10.4 & 0.1 & 1.6 &
1.72 & 2.78 & $-$0.01 & $-$ \\
RS 8     & 13.8 & $-0.09 \pm 0.02$ & $-0.22 \pm 0.02$ &  5.8 & 0.3 & 2.5 &
1.56 & 2.79 &    0.15 & $-$ \\
RS 72    & 13.9 & $-0.10 \pm 0.02$ & $-0.18 \pm 0.02$ &  5.5 & 0.3 & 2.5 &
1.71 & 2.81 & $-$0.01 & $-$ \\
RS 162   & 14.1 & $-0.03 \pm 0.01$ & $-0.13 \pm 0.02$ &  5.9 & 0.3 & 1.9 &
1.66 & 2.81 &    0.04 & $-$ \\
BO 24    & 14.3 & $-0.06 \pm 0.03$ & $-0.19 \pm 0.03$ &  3.6 & 0.4 & 2.9 &
1.72 & 3.00 & $-$0.04 & $-$0.24 \\
BO 26    & 14.6 & $-0.13 \pm 0.03$ & $-0.21 \pm 0.02$ &  3.0 & 0.5 & 2.8 &
1.70 & 2.70 & $-$0.02 & $-$ \\
BO 25    & 14.6 & $-0.04 \pm 0.02$ & $-0.06 \pm 0.02$ &  5.3 & 0.3 & 1.8 &
1.55 & 2.78 &    0.12 & $-$ \\
BO 34    & 15.0 & $-0.08 \pm 0.03$ & $-0.03 \pm 0.09$ &  6.4 & 0.2 & 1.3 &
1.53 & 2.69 &    0.12 & $-$ \\
\hline
\end{tabular}
 \caption{List of the galaxies with $K$ band photometry. Col. (2), (3),
 and (5): From Paper I. Effective radii ($r_e$) are estimated in the
 $R-$band. Col. (6) and (7): The inner and outer cutoff radii are
 indicated in unit of effective radius. Col. (8) and (9): $B-R$ and
 $R-K$ colours estimated within an aperture of 12$^{\prime\prime}$ (5
 $h^{-1}$ kpc) in diameter, which is equivalent to that in Bower et
 al. (1992) for Coma ellipticals.  Col. (10) and (11): The amounts of
 the offsets applied to $B-R$ and $R-K$ colours, respectively.}
 \label{list1}
\end{minipage}
\end{table*}

\subsection{The $K$ band observation}

The $K$ band imaging observation was performed on 20 and 21 June 2003
with UFTI on UKIRT 3.8m. UFTI has a 1K$\times$1K array and one pixel of
the detector subtends $0_{\cdot}^{\prime\prime}091$ on sky, yielding a
field of view of $1_{\cdot}^{\prime}5 \times 1_{\cdot}^{\prime}5$. A
typical seeing size during the observing run was $\sim
0_{\cdot}^{\prime\prime}7$ in FWHM. An exposure time of each frame was
40 sec or 60 sec and a total on-source integration time is typically
1,200 sec with the dither pattern of 9 telescope pointings. Since the
apparent size of a target is too extended to make flat-field frames from
object frames themselves, the on-source exposures were interlaced by
exposures of adjacent blank sky regions with the same exposure time as
that of the object frames.

These $K$ band imaging data were reduced with IRAF. After dark
subtraction, we carried out flat-fielding using sky-flat frames. To make
a sky-flat frame, we stacked $\sim$ 5 blank sky frames and normalised
the stacked image by its modal pixel value. An object frame was
flat-fielded by the sky-flat frame constructed with the blank sky frames
which were taken close in observation time to the object frame. This
method allows us to correct for the complex pattern seen in the
flat-field frame whose contrast slightly varies with time. The
flat-fielded blank sky frames have very uniform distribution of
brightness across the field of view (global variation is $\sim$ 0.1 \%
or even smaller and pixel-to-pixel variation is $\sim$ 1 \%), which is
reasonable for an image looking at such a small area on the sky.

Sky subtraction was performed by estimating a single background value in
each object frame, where the following two steps were employed. Firstly,
a sky background in each frame was estimated by calculating the modal
pixel value after bright objects were masked and was subtracted.  The
images taken with the telescope dithered around each target were
registered by sub-pixel shifts and these registered images were stacked
with the 3 $\sigma$ clipping algorithm.  Secondly, since objects which
are too faint to be found in the individual exposure but may cause a
wrong estimation of sky background are detectable in this stacked image,
they were picked out using SExtractor version 2.2.2 (Bertin \& Arnout
1996). The coordinates of these objects were de-registered to those in
the unregistered frames and the pixels where the objects should be were
masked. The sizes and shapes of the masks were defined based on those
measured in the stacked image.  Then, a sky background was re-estimated
in each of the masked images and it was subtracted.

These sky-subtracted images were registered by sub-pixel shifts and,
after the FWHMs of stellar objects in the sky-subtracted and registered
frames were matched by convolutions of images with Gaussian kernels,
they were stacked with the 3 $\sigma$ clipping algorithm. Finally, the
$K$ band stacked image was aligned with the optical data by using stars
in the field of view and a Gaussian convolution was employed to match
the FWHMs of stellar objects with those in the optical images.

\subsection{Calibration}

For the optical data, the photometric calibration was performed using
the standard stars in Landolt (1992). Several standard star fields were
observed at the beginning and end of each night. Since the weather was
slightly non-photometric, an external check of the accuracy of the
calibration was performed using aperture photometry data and growth
curves for several galaxies, which were taken from
HYPERCAT\footnote{http://www-obs.univ-lyon1.fr/hypercat}. There seem to
be zero-point offsets between our growth curves and those from HYPERCAT;
the amounts of the offsets are $0 - 0.1$ mag in each band. We will come
back to the treatment of these offsets later.

For the $K$ band data, the photometric calibration was carried out using
UKIRT Faint Standard stars (Hawarden et al. 2001).  About 10 standard
stars were observed at the beginning and end of each night. The weather
was clear and stable during the two nights
and the photometric zero-point can be determined with an accuracy of
$\sim$ 0.02 mag from the data of standard stars.

It is found that some of the galaxies have $B-R$ colours slightly
inconsistent ($\sim 0.1$ mag) with those of typical luminous ellipticals
in the local universe and this can be due to the calibration error in
$B$ and $R$ bands. We correct for these offsets using the CM relation.
We assume that the CM relation in the Coma cluster also exists in ABELL
2199, and their ``true'' colours are estimated using stellar populations
modelled by Kodama et al. (1998) using a population synthesis model so
that the CM relations in $U-V$ and $V-K$ colours obtained in the Coma
cluster by Bower et al. (1992) are well reproduced. We estimated $B-R$
and $R-K$ colours of the sample galaxies within an aperture of
12$^{\prime\prime}$ (5 $h^{-1}$kpc) in diameter (Table \ref{list1}),
which is equivalent to that in Bower et al. (1992), and estimated the
deviations of the colours from the CM relations at their luminosities
and adopted the values as zero-point offsets of the colours. The amount
of the offset applied to each object is listed in Table
\ref{list1}. These are corrected for in the following presentations of
the colours.
We note that the $R-K$ colour of BO 24 is $\sim$ 0.2 mag redder than the
others. Since its $B-R$ colour is not unusual, we also attribute this
shift to an observational error and corrected for it using the CM
relation in the same method as above.
It should be stressed that these corrections are performed by shifting
zero points of colours and do not affect any spatial variations of
colour in galaxies, which are focused on in this paper. We performed
these corrections only for clarity in comparing the data with variations
of colours predicted by population synthesis models as described in the
next section.

\section{DATA ANALYSES AND RESULTS}

\subsection{Radial Profiles of Surface Brightness and Colours in Galaxies}

\begin{figure*}
 \centering
 \includegraphics[width=13cm,keepaspectratio,clip]{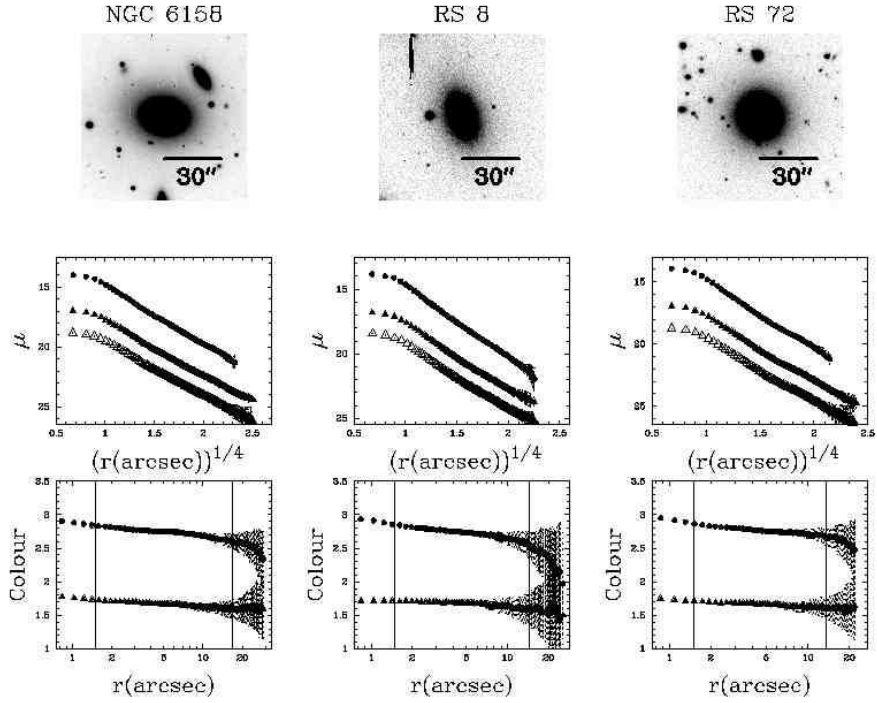}
 \caption{$R$ band image (top), radial profiles of surface brightness
 (middle), and colour (bottom). In the middle panels, open triangles,
 solid triangles, and circles indicate $B$, $R$, and $K$ band surface
 brightness profiles, respectively. In the bottom panels, triangles and
 circles indicate $B-R$ and $R-K$ colours, respectively, and two vertical
 lines show the inner and outer cutoff radii.} \label{profs}
\end{figure*}
\setcounter{figure}{1}
\begin{figure*}
 \centering
 \includegraphics[width=13cm,keepaspectratio,clip]{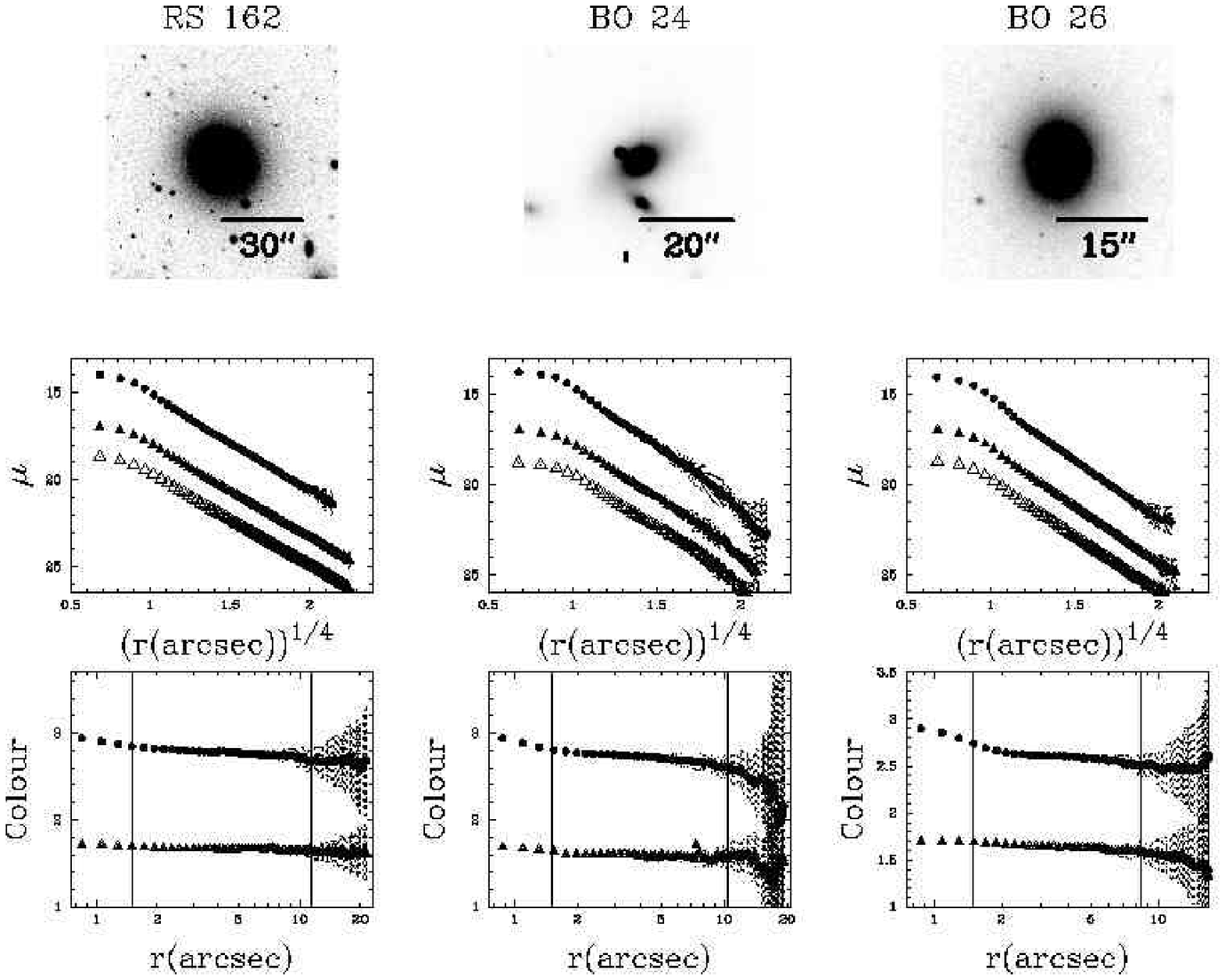}
 \caption{Radial profiles of surface brightness and colour --
 continued.}
\end{figure*}

\setcounter{figure}{1}
\begin{figure*}
 \centering
 \includegraphics[width=13.5cm,keepaspectratio,clip]{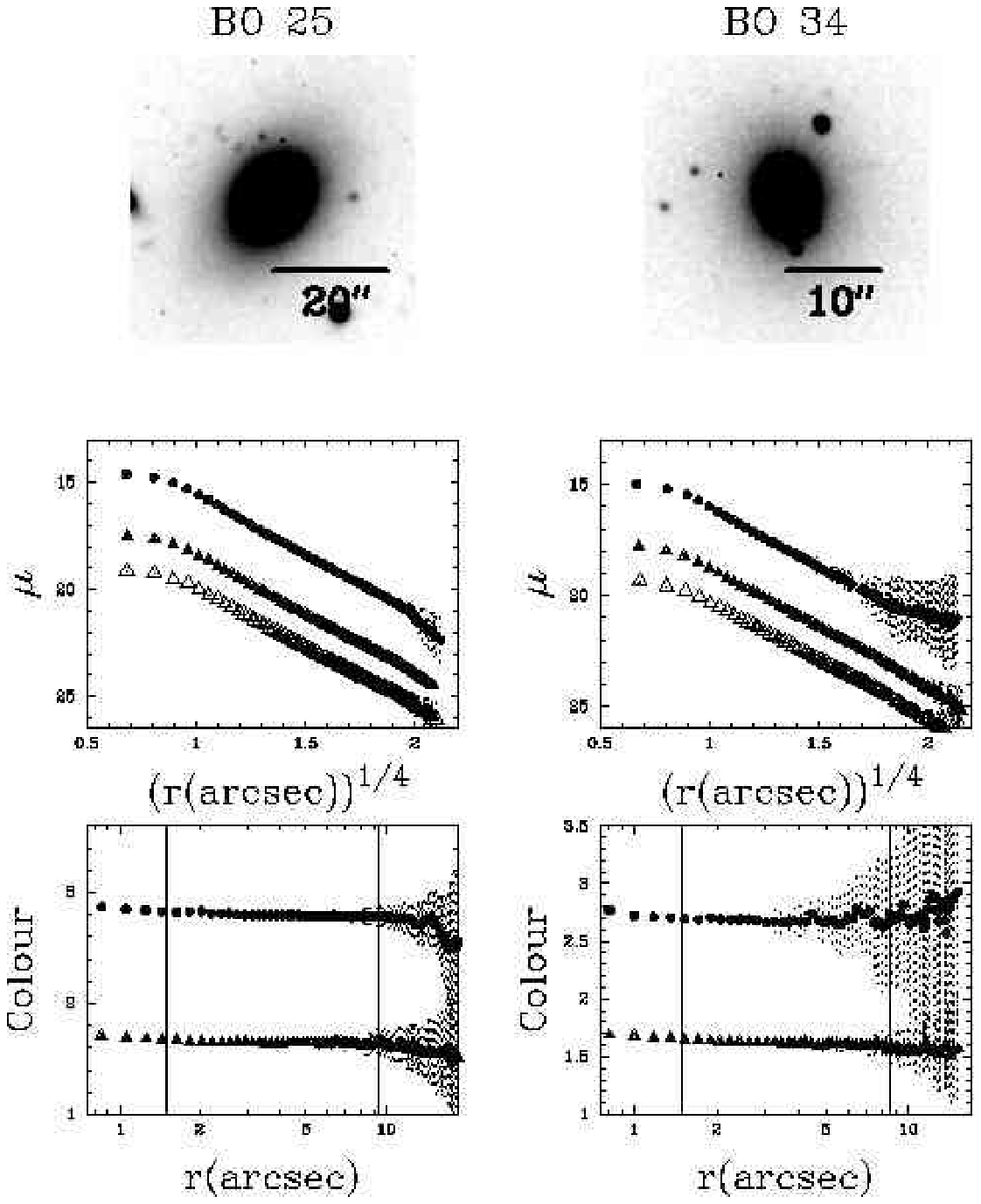}
 \caption{Radial profiles of surface brightness and colour --
 continued.}
\end{figure*}

\setcounter{figure}{2}
\begin{figure*}
 \centering
 \includegraphics[height=16cm,angle=-90,keepaspectratio,clip]{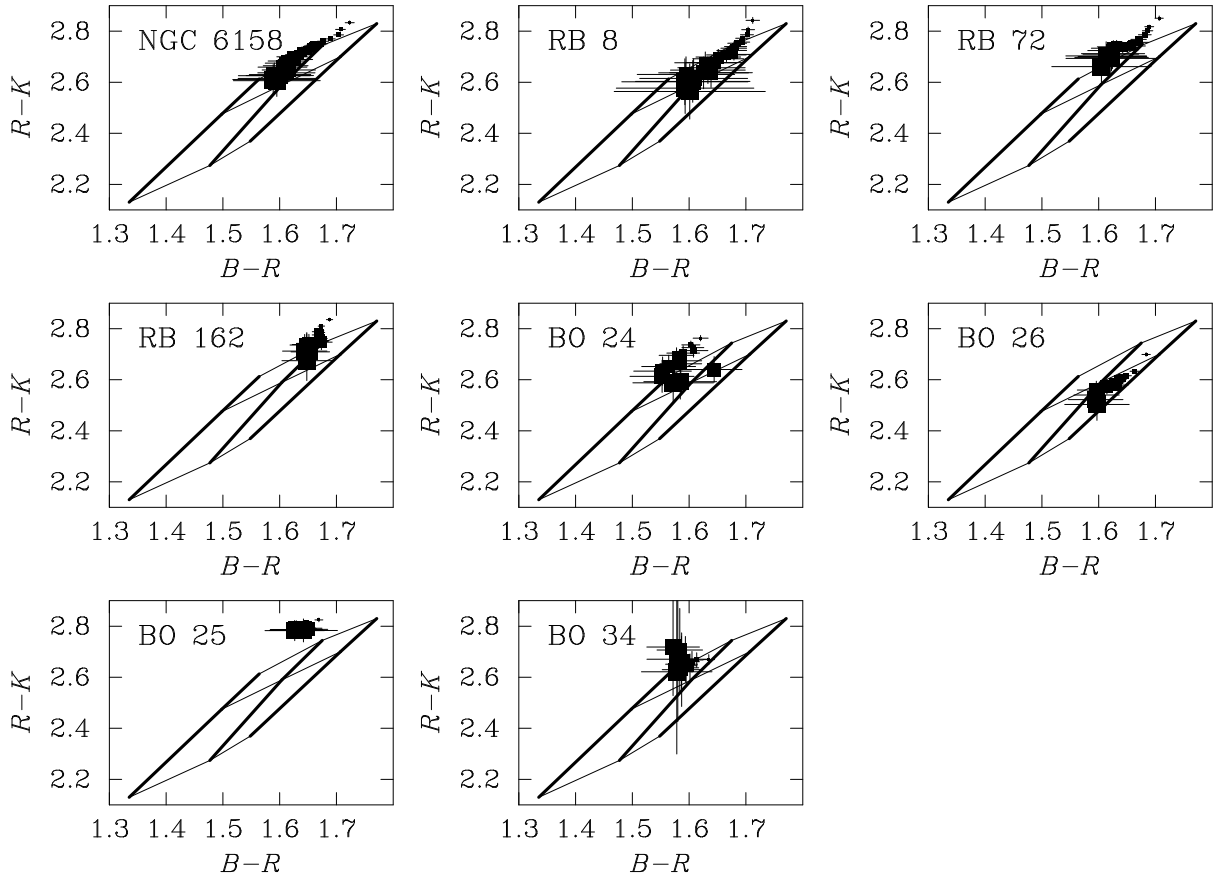}
 \caption{$B-R$ and $R-K$ colour-colour diagrams. The galaxies are
 arranged in order of decreasing luminosity from left to right and top
 to bottom right (NGC 6158 is the most luminous and BO 34 is the least
 luminous). The grid superposed with the data points on each diagram
 shows variations of $B-R$ and $R-K$ colours of an SSP along with its
 age and metallicity predicted using PEGASE for the Salpeter IMF. Bold
 line indicates colour variation predicted for a constant age, and thin
 line depicts that for a constant metallicity. More details of the grid
 are explained in Figure \ref{grid}. Note that the same ranges of
 colours are shown in all panels.} \label{twocol}

\end{figure*}

The radial surface brightness profiles of the galaxies in $B$, $R$, and
$K$ bands are indicated in Figure \ref{profs}. These profiles were
obtained along ellipses fitted to the isophotes with the ELLIPSE task in
the STSDAS package with a radial sampling of
0$_{\cdot}^{\prime\prime}$22 (1 pix) along major axis. Since bright
objects in the neighbor of a target galaxy may have serious effects on
the fitted ellipses, the neighbors are masked out beforehand. In fitting
an ellipse to an isophote, ellipticity is set to be a free parameter.
Galaxy centre is fixed to a centroid of a galaxy in the $R$ band image
and position angle is also fixed to that of an ellipse within which half
of the galaxy light is included. In the following analyses, all of the
radial profiles and the parameters related to radius are expressed in
terms of the equivalent radius of an ellipse: $\sqrt{ab}$, where $a$ and
$b$ are the semi-major and semi-minor axes of the ellipse.

Before obtaining radial colour profiles in the sample galaxies, we made
the $B-R$ colour maps of the galaxies after the seeing sizes in both of
$B$ and $R$ bands were matched (Paper I). They demonstrate that almost
all the galaxies show axisymmetric colour distributions independent of
morphology and luminosity and that their radial colour profiles and the
colour gradients well represent the two dimensional colour
distributions. We also made the $R-K$ colour maps of the galaxies
observed in the $K$ band and found that their trends are the same as
those in $B-R$.

The radial profiles of $B-R$ and $R-K$ colour of a sample galaxy are
constructed by subtracting the surface brightness profile in the $R$
band from those in $B$ and $K$ bands, both of which were made along the
same ellipses fitted to isophotes in the $R$ band image. The results are
shown in the bottom panels of Figure \ref{profs}. An error bar attached
to each data point in the colour profile includes a local sky
subtraction error and a standard deviation of colours along each
elliptical isophote.
To estimate the sky subtraction error, we construct a histogram of pixel
values in blank regions of a fully reduced image where surface
photometry of a galaxy is performed. Then we adopt the standard
deviation around the modal value as an error of background estimation.
In the optical data, the pixel values in an annulus around each target
whose inner radius and width are $\sim 3R_{25}$ and $11^{\prime\prime}$
(50 pix), respectively, were used to make the histogram. In the $K$ band
data, pixels which remained after objects were masked were
investigated in each object frame.
In the following, we focus on the portions of the galaxies between the
inner and outer cutoff radii, which were defined to avoid the regions
where colour distributions are seriously affected by seeing effects and
those where colours are rather poorly determined.

The outer cutoff radius is defined to be a radius where the $R$ band
surface brightness of a galaxy is 22.5 mag/arcsec$^2$ (Paper I). At this
radius, the typical error of $B-R$ colour amounts to $\sim$ 0.1 mag.
Since the typical error of $R-K$ colour is also $\sim$ 0.1 mag at this
radius, it is reasonable to adopt the same cutoff as that applied to the
optical data in investigating the radial profiles of $R-K$ colour and
estimating the gradients. The $R-K$ colour profile of BO 34 (Figure
\ref{profs}) starts becoming noisy at a smaller radius than in the other
galaxies due to the $\sim$ 40 \% shorter integration time in the $K$
band.

The inner cutoff radius is necessary to mitigate seeing effects on
colour gradients. In order to define this, we performed simulations
using artificial elliptical galaxies added on the real images to
investigate deviations of measured colour gradients from the intrinsic
values caused by seeing effects. We also investigated their dependences
on size, luminosity, ellipticity, and intrinsic colour gradient of model
elliptical galaxy (see Paper I for details). Based on the simulations,
we defined the inner cutoff radius to be $1_{\cdot}^{\prime\prime}5$,
which can reduce the deviation of a measured colour gradient from the
intrinsic value down to the level comparable to a fitting error of the
regression line.
It should be mentioned that since the FWHMs of stellar objects in the
$K$ band images are smaller than those in the $B$ and $R$ band images
and the $K$ band data were smoothed to match the FWHMs with those in the
optical data, this inner cutoff is not required to be modified.

To summarise, the same inner and outer cutoffs defined in $B-R$ colour
can be adopted to the radial colour profiles in $R-K$. In Table
\ref{list1}, the cutoff radii for each galaxy scaled by the effective
radius are tabulated. The $B-R$ and $R-K$ colour gradients calculated by
fitting regression lines to the colour profiles between the inner and
outer cutoff radii are also listed.

\subsection{Spatially Resolved $B-R$ and $R-K$ Diagrams}

In Figure \ref{twocol}, the $B-R$ and $R-K$ colours of each target
galaxy within the region between the inner and outer cutoff radii are
shown. The colours at outer radii are indicated with larger symbols and
the radial sampling is the same as that in the radial colour profiles in
Figure \ref{profs} (0$_{\cdot}^{\prime\prime}$22). The two colour
diagrams are ordered in decreasing luminosity from left to right and top
to bottom (NGC 6158 is the most luminous and BO 34 is the least
luminous). The data points in each panel show a basic trend of radial
colour variation that colours are redder in both of $B-R$ and $R-K$ at
smaller radii. It is interesting to note that although the sample is
small, there seems to be a variety in distribution of the colours on the
diagram from galaxy to galaxy.

\begin{figure}
 \centering
 \includegraphics[height=7cm,angle=-90,keepaspectratio,clip]{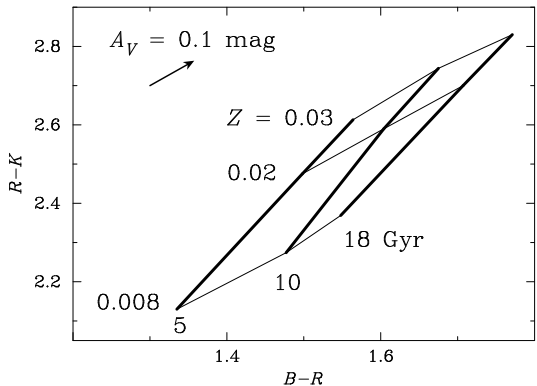}
 \caption{Details of the SSP model grid with PEGASE superposed with the
 data in Figure \ref{twocol}. The reddening vector for $A_V = 0.1$ mag
 is also shown for comparison.} \label{grid}
\end{figure}

Although it is hard to disentangle the age-metallicity degeneracy, it is
still possible to obtain some insight into how stellar age and
metallicity vary across a galaxy by comparing the $B-R$ and $R-K$ colour
variations with predictions from a simple stellar population (SSP)
model. The grid superposed on the two colour diagrams indicates
variations of $B-R$ and $R-K$ colours of an SSP model along with its age
and metallicity. This is constructed using the SSP models by PEGASE Ver
2.0 (Fioc \& Rocca-Volmerange 1997) with the Salpeter Initial Mass
Function (IMF). The details of the SSP models for the grid are indicated
in Figure \ref{grid}.
We sample the ages of 5, 10, and 18 Gyr and the metallicities of 0.008
(0.4 $Z_{\odot}$), 0.02 ($Z_{\odot}$), and 0.03 (1.5 $Z_{\odot}$). It is
stressed that only relative values of age and metallicity from a model
are discussed here and determining their absolute values in a galaxy is
beyond the scope of this paper.

Comparing the data points with the grid in Figure \ref{twocol}, it is
suggested that the colour distributions tend to be in parallel with the
prediction for a constant age and it seems to be dominated by a
metallicity gradient in the sense that stars are more metal rich in the
inner regions. In most of the ellipticals, the data do not seem to
support that the primary origin of the colour gradients is an age
gradient.

\begin{figure*}
 \centering
 \includegraphics[height=17cm,angle=-90,keepaspectratio,clip]{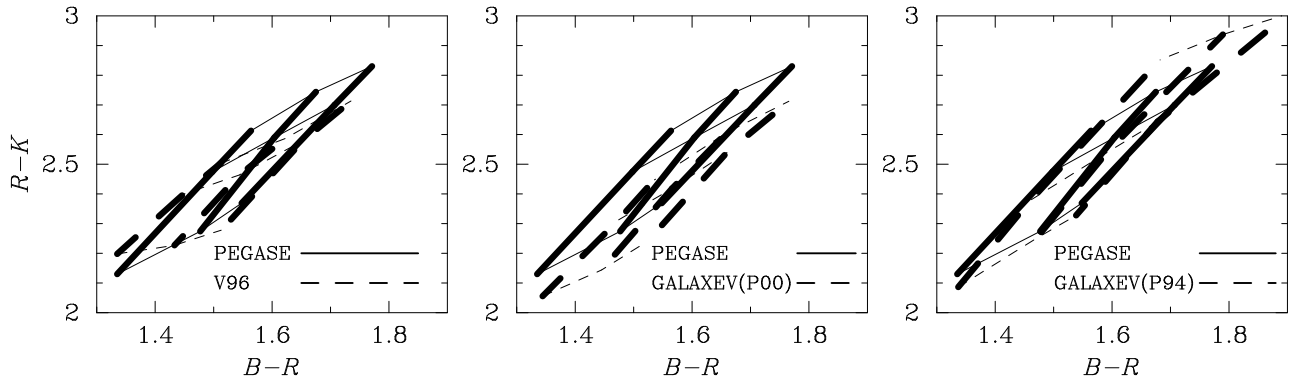}
 \caption{Grids constructed with different SSP models are shown. In each
 panel, the colour variation predicted with PEGASE (solid line) is
 compared with that predicted with V96 or GALAXEV (dashed line). The
 Salpeter IMF is adopted in all the models. For the grid from GALAXEV
 with the P94 tracks (right panel), the ages of 5, 10, and 18 Gyr and
 and the metallicities of $Z = 0.008, 0.02$, and 0.05 are used. For the
 other grids, the ranges of age and metallicity are the same as those in
 Figure \ref{grid}; 5, 10, and 18 Gyr in age and $Z = 0.008, 0.02$, and
 0.03 in metallicity. Note that slightly larger ranges of colours than
 those in Figure \ref{twocol} are used for the grids to be included
 entirely.} \label{gridcomp1}

\end{figure*}

\begin{figure}
 \centering
 \includegraphics[width=7cm,keepaspectratio,clip]{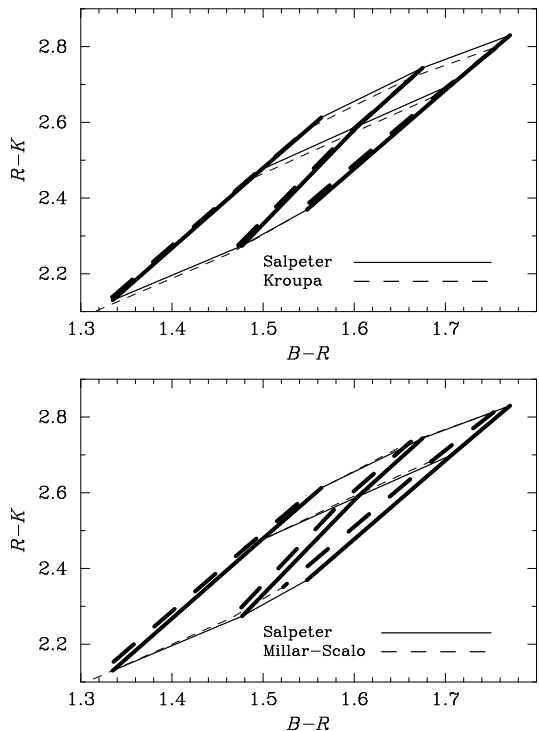}
 \caption{Variation of the grid due to choice of IMF is exemplified. In
 each panel, the colour variation predicted with the Salpeter IMF
 (solid line) is compared with that with the Kroupa IMF or the
 Millar-Scalo IMF (dashed line). All the grids are constructed using
 PEGASE.} \label{gridcomp2}
\end{figure}

It is worth investigating how much the grid on the two colour diagram
can vary from model to model. We compare the grid by PEGASE with that by
Vazdekis et al. (1996; hereafter V96) and those by GALAXEV (Bruzual \&
Charlot 2003) with the Padova 1994 (P94) or Padova 2000 (P00)
evolutionary tracks in Figure \ref{gridcomp1} where the Salpeter IMF is
adopted for all the models. The lower and upper mass cutoffs are 0.1 and
100 $M_{\odot}$ in PEGASE, 0.0992 $M_{\odot}$ and 72 $M_{\odot}$ in V96,
and 0.1 $M_{\odot}$ and 120 $M_{\odot}$ in GALAXEV. Note that for the
GALAXEV model with the P94 evolutionary tracks, the metallicity of the
most metal rich SSPs is 2.5 times larger than the solar value, while it
is 1.5 times larger than the solar value for all the other models. By
using PEGASE, variation of the grid due to choice of IMF is also
investigated. The grids for three representative IMFs are indicated in
Figure \ref{gridcomp2}; the Salpeter IMF, the Kroupa IMF (Kroupa, Tout
\& Gilmore 1993), and the Millar-Scalo IMF (Miller \& Scalo 1979). As
shown in Figure \ref{gridcomp1}, there is some discrepancy in the shape
of the grid as well as its location on the two colour diagram among the
models. On the other hand, Figure \ref{gridcomp2} suggests that both of
the shape and location of the grid are rather insensitive to choice of
IMF.

Considering contributions of both age and metallicity gradients, a
colour gradient can be described as follows:
\begin{equation}
 {{d (colour)}\over{d \log r}} = a~{{d \log Z}\over{d \log r}} + b~{{d \log {\rm (age)}}\over{d \log r}},
\end{equation}
where
\begin{equation}
a \equiv {{\partial (colour)}\over{\partial \log Z}}\Bigr|_{\rm age},
\end{equation}
and
\begin{equation}
b \equiv {{\partial (colour)}\over{\partial \log {\rm (age)}}}\Bigr|_Z.
\end{equation}
Given the sensitivities of $B-R$ and $R-K$ colours to age and
metallicity (i.e., $a$ and $b$) which can be estimated with an SSP
model, the age and metallicity gradients can be worked out
simultaneously using the $B-R$ and $R-K$ colour gradients derived with
the data.
A similar method has been applied to absorption line strength gradients
(Kobayashi \& Arimoto 1999; Mehlert et al. 2003; see also Henry \&
Worthey 1999).
The coefficient of $a$ is determined within a range of metallicity ($0.4
< Z/Z_{\odot} < 1.5$) by using colours predicted with SSPs having a
single age (10 Gyr). The coefficient of $b$ is calculated within a range
of age between 5 Gyr and 18 Gyr by using SSPs having the solar
metallicity.  We calculated these coefficients for each of the SSP
models (PEGASE, V96, and GALAXEV with P94\footnote{For GALAXEV with P94,
the coefficient $a$ is determined for $0.4 < Z/Z_{\odot} < 2.5$.} and
P00) with the Salpeter IMF assumed (Table \ref{sensitivities}) and
derived age and metallicity gradients of a galaxy. The results from our
data are shown in Table \ref{list2}, indicating that negative
metallicity gradients (stellar metallicity is higher towards the galaxy
centre) are detected in all the cluster ellipticals but BO 34, while an
age gradient is significantly detected only in NGC 6158.
This suggests that the cluster ellipticals have metallicity gradients
while their age gradients are consistent with zero; using PEGASE, the
typical metallicity gradient is estimated to be $-0.16 \pm 0.09$ in $d
\log Z /d \log r$, while the age gradient is estimated to be $-0.10 \pm
0.14$ in $d \log {\rm (age)} /d \log r$.
It is mentioned that although estimated values of age and metallicity
gradients depend on model, the qualitative trend that the cluster
ellipticals have negative metallicity gradients but age gradients are
not detected significantly can be seen in all the models.

\setlength{\tabcolsep}{3mm}
\begin{table}
 \centering
 \begin{tabular}{lcccc}
 \hline
 \multicolumn{1}{c}{Model} & \multicolumn{2}{c}{$B-R$} &
 \multicolumn{2}{c}{$R-K$} \\ \cline{2-3} \cline{4-5}
     & $a$ & $b$ & $a$ & $b$ \\
 \multicolumn{1}{c}{(1)}   & (2) & (3) & (4) & (5) \\ \hline
 PEGASE        & 0.34 & 0.38 & 0.82 & 0.40 \\
 V96           & 0.33 & 0.37 & 0.63 & 0.32 \\
 GALAXEV (P94) & 0.56 & 0.39 & 1.20 & 0.47 \\
 GALAXEV (P00) & 0.27 & 0.33 & 0.58 & 0.41 \\ \hline
 \end{tabular}
 \caption{Sensitivities of $B-R$ and $R-K$ colours to age and
  metallicity. The coefficient $a$ ($\equiv (\partial (colour) /
  \partial \log Z)_{\rm age}$) is determined within a range of
  metallicity ($0.4 < Z/Z_{\odot} < 2.5 $ for GALAXEV with P94 and $0.4
  < Z/Z_{\odot} < 1.5$ for the other models) by using SSPs with a 10 Gyr
  age. The coefficient $b$ ($\equiv (\partial (colour) / \partial \log
  {\rm age})_Z$) is determined within a range of age between 5 Gyr and
  18 Gyr by using SSPs with the solar metallicity.}
  \label{sensitivities}
\end{table}

\section{DISCUSSION}

\setlength{\tabcolsep}{3mm}
\begin{table*}
 \centering
 \begin{minipage}{105mm}
 \caption{Gradients of age and metallicity estimated from the $B-R$
 and $R-K$ gradients.} \label{list2}
 \begin{tabular}{llcc}
 \hline
 Galaxy & \multicolumn{1}{c}{Model} &
 $d \log {\rm (age)}/ d \log r$ & $d \log Z / d \log r$
 \\		 
        &             &
 (dex$^{-1}$)                   & (dex$^{-1}$)
 \\		 
 \multicolumn{1}{c}{(1)} & \multicolumn{1}{c}{(2)} &
 (3)                            & (4)
 \\
 \hline
         & PEGASE        & $-0.24 \pm 0.07$ & $-0.09 \pm 0.05$ \\
NGC 6158 & V96           & $-0.15 \pm 0.08$ & $-0.19 \pm 0.06$ \\
         & GALAXEV (P94) & $-0.24 \pm 0.09$ & $-0.05 \pm 0.04$ \\
         & GALAXEV (P00) & $-0.28 \pm 0.10$ & $-0.09 \pm 0.09$ \\ \hline
         & PEGASE        & $+0.01 \pm 0.14$ & $-0.28 \pm 0.09$ \\
RS 8     & V96           & $+0.13 \pm 0.16$ & $-0.42 \pm 0.11$ \\
         & GALAXEV (P94) & $+0.07 \pm 0.17$ & $-0.21 \pm 0.09$ \\
         & GALAXEV (P00) & $+0.08 \pm 0.20$ & $-0.44 \pm 0.18$ \\ \hline
         & PEGASE        & $-0.12 \pm 0.14$ & $-0.16 \pm 0.09$ \\
RS 72    & V96           & $-0.03 \pm 0.16$ & $-0.27 \pm 0.11$ \\
         & GALAXEV (P94) & $-0.10 \pm 0.17$ & $-0.11 \pm 0.09$ \\
         & GALAXEV (P00) & $-0.11 \pm 0.20$ & $-0.23 \pm 0.18$ \\ \hline
         & PEGASE        & $+0.12 \pm 0.09$ & $-0.22 \pm 0.07$ \\
RS 162   & V96           & $+0.20 \pm 0.10$ & $-0.31 \pm 0.09$ \\
         & GALAXEV (P94) & $+0.18 \pm 0.11$ & $-0.18 \pm 0.06$ \\
         & GALAXEV (P00) & $+0.21 \pm 0.13$ & $-0.37 \pm 0.13$ \\ \hline
         & PEGASE        & $+0.10 \pm 0.21$ & $-0.28 \pm 0.14$ \\
BO 24    & V96           & $+0.20 \pm 0.23$ & $-0.41 \pm 0.17$ \\
         & GALAXEV (P94) & $+0.17 \pm 0.26$ & $-0.23 \pm 0.13$ \\
         & GALAXEV (P00) & $+0.19 \pm 0.30$ & $-0.46 \pm 0.26$ \\ \hline
         & PEGASE        & $-0.20 \pm 0.19$ & $-0.16 \pm 0.12$ \\
BO 26    & V96           & $-0.10 \pm 0.21$ & $-0.28 \pm 0.14$ \\
         & GALAXEV (P94) & $-0.19 \pm 0.23$ & $-0.10 \pm 0.11$ \\
         & GALAXEV (P00) & $-0.23 \pm 0.27$ & $-0.20 \pm 0.22$ \\ \hline
         & PEGASE        & $-0.07 \pm 0.14$ & $-0.04 \pm 0.09$ \\
BO 25    & V96           & $-0.04 \pm 0.16$ & $-0.07 \pm 0.11$ \\
         & GALAXEV (P94) & $-0.07 \pm 0.17$ & $-0.02 \pm 0.09$ \\
         & GALAXEV (P00) & $-0.08 \pm 0.20$ & $-0.04 \pm 0.18$ \\ \hline
         & PEGASE        & $-0.33 \pm 0.33$ & $+0.13 \pm 0.27$ \\
BO 34    & V96           & $-0.33 \pm 0.39$ & $+0.12 \pm 0.35$ \\
         & GALAXEV (P94) & $-0.39 \pm 0.43$ & $+0.13 \pm 0.24$ \\
         & GALAXEV (P00) & $-0.46 \pm 0.49$ & $+0.27 \pm 0.50$ \\
\hline
\end{tabular}
\end{minipage}
\end{table*}

It has also been suggested from the other recent studies that nearby
ellipticals have negative metallicity gradients while their age
gradients are consistent with zero (Mehlert et al. 2003; Wu et al.
2004). The presence of a metallicity gradient and the absence of an age
gradient are consistent with the monolithic collapse scenario. The
typical metallicity gradient estimated from gradients of colours and
metal absorption line strengths in nearby ellipticals could be
reproduced in a recent model of galaxy formation through a
monolithic-like process (e.g., Kawata 1999). Since all stars are
presumed to form nearly coevally at high redshift in this scenario, no
significant variation in stellar age across a galaxy would be expected.
On the other hand, it is less clear what kind of radial variations of
stellar populations can be formed in elliptical galaxies in the
framework of hierarchical galaxy formation, while it has recently been
suggested by numerical simulations that metallicity gradients estimated
in nearby ellipticals assuming no age gradients can be reproduced
including its variety (Kobayashi 2004).
It is intriguing to mention that unlike the monolithic collapse
scenario, an age gradient can be acquired due to star formation
associated with mergers and/or accretions. Exploring what kind of age
gradients elliptical galaxies could possess in the hierarchical galaxy
formation may provide clues to understanding their formation processes
and evolutionary histories (Benson, Ellis \& Menanteau 2002).

Our data also suggest that colour distributions on the two colour
diagram tend to be less scattered in less luminous cluster ellipticals.
In other words, stellar populations may be spatially more uniform in
less luminous cluster ellipticals. This is not because in less luminous
ellipticals colours are investigated within the smaller portions of
galaxies; the outer cutoff radii scaled by effective radii are similar
in all the galaxies (Table \ref{list1}).
If the colour gradients originate from pure metallicity gradients, this
trend would indicate that the metallicity gradients correlate with the
galaxy luminosities and could support the monolithic collapse scenario
(Larson 1974; Carlberg 1984; Kawata \& Gibson 2003). However, the
metallicity gradients estimated using both of the $B-R$ and $R-K$ colour
gradients do not seem to well correlate with the luminosities.
This may simply be due to the poor statistics and/or the individual
measurements of the metallicity gradients may be too coarse.
More precise estimations of age and metallicity gradients for a larger
sample of galaxies will be necessary for detailed studies of the
relationship between radial variation of stellar population and galaxy
luminosity in the cluster ellipticals.

Age and metallicity gradients in nearby ellipticals have recently been
investigated in a range of galaxy environment. In Mehlert et al. (2003),
the age and metallicity gradients in 35 elliptical and S0 galaxies in
the Coma cluster were estimated from the gradients of absorption line
indices (H$\beta$, Mg$b$ and $<$Fe$>$). In Wu et al. (2004), the
gradients in 36 elliptical and S0 galaxies in the Sloan Digital Sky
Survey Early Data Release which are sampled not from a specific cluster
but from various environments were estimated with multi band surface
photometry from optical to NIR. Including our study of the luminous
early-type galaxies in a rich cluster ABELL 2199, the presence of a
metallicity gradient and the absence of an age gradient have been
suggested in all of these studies. This may imply that there is no
strong dependence on galaxy environment in radial gradient of stellar
population in elliptical galaxy.

Finally, it should be mentioned that dust extinction may have some
effects on a colour gradient in an elliptical galaxy (e.g., Goudfrooij
\& de Jong 1995). As a matter of fact, even if an elliptical galaxy
consists of a mixture of stars without any radial gradients of stellar
population and diffusely distributed dust, a calculation of the
radiative transfer in the galaxy suggests that the colour gradients
could be reproduced only by the dust effects (Witt, Thronson, \& Capuano
1992; Wise \& Silva 1996). However, many elliptical galaxies show not
only colour gradients but also metal absorption line strength gradients,
which are unlikely to be created by effects of dust extinction. It has
also been suggested that {\it on average}, metallicity gradients in
elliptical galaxies as estimated by a population synthesis model from
colour gradients are consistent with those estimated from absorption
line index gradients (e.g., Peletier et al. 1990; Davies et al. 1993).
Nevertheless there are some exceptions and it is hard to isolate the
effect of dust extinction on an individual galaxy basis. Future
observation in the far-infrared with high spatial resolution may be able
to give constraints on such a spatial variation of effects of dust
extinction.

\section{SUMMARY}

We performed $K$ band surface photometry for luminous early-type
galaxies in a nearby rich cluster ABELL 2199. Combining it with $B$ and
$R$ band surface photometry, radial variations of $B-R$ and $R-K$
colours in the galaxies were investigated. It is found that the inner
regions of the galaxies are redder in both of $B-R$ and $R-K$ colours.
Comparing the radial variations of both of the colours with predictions
of SSP models for a range of ages and metallicities, it is suggested
that cluster ellipticals have metallicity gradients but their age
gradients are consistent with zero, although the sample is small; the
typical metallicity gradient is estimated to be $-0.16 \pm 0.09$ in $d
\log Z /d \log r$, while the age gradient is estimated to be $-0.10 \pm
0.14$ in $d \log {\rm (age)} /d \log r$.  Since similar results have
also been obtained in the other recent studies by investigating
ellipticals in the Coma cluster and less dense environments, it seems
that radial gradients of stellar populations in elliptical galaxies and
thus their evolutionary histories are rather insensitive to galaxy
environment.

Considering the trend found in our optical study that less luminous
ellipticals have less steep colour gradients, it is suggested that they
have spatially more uniform distributions of both $B-R$ and $R-K$
colours and thus probably stellar population.
However the metallicity gradients estimated using both of the $B-R$ and
$R-K$ colours do not seem to well correlate with the galaxy
luminosities. This may simply be due to the poor statistics and/or the
individual measurements of the metallicity gradients may be too coarse.
More precise estimations of age and metallicity gradients for a larger
sample of galaxies will be necessary for detailed studies of the
relationship between radial variation of stellar population and galaxy
luminosity in the cluster ellipticals.

\section*{Acknowledgments}

We are grateful to the staff of the Joint Astronomy Centre for its
support of the UKIRT observation. We appreciate the support from the
members of the University of Hawaii observatory. We also thank the
referee, Dr. Francesco La Barbera, for helpful comments and advice. This
research made use of the NASA/IPAC Extragalactic Database (NED), which
is operated by the Jet Propulsion Laboratory, California Institute of
Technology, under a contract with the National Aeronautics and Space
Administration.

\label{lastpage}


\begin{thebibliography}{}
\bibitem[\protect\citeauthoryear{Benson, Ellis \& Menanteau}{2002}]{benson}
Benson, A. J., Ellis, R. S., \& Menanteau, F. 2002, MNRAS, 336, 564

\bibitem[\protect\citeauthoryear{Bertin \& Arnouts}{1996}]{sext}
Bertin, E., \& Arnouts, S. 1996, A\&AS, 117, 393

\bibitem[\protect\citeauthoryear{Bower, Lucey \& Ellis}{1992}]{bower}
Bower, R. G., Lucey, J. R., \& Ellis, R. S. 1992, MNRAS, 254, 601

\bibitem[\protect\citeauthoryear{Bruzual \& Charlot}{2003}]{bc}
Bruzual, G., \& Charlot, S., 2003, MNRAS, 344, 1000

\bibitem[\protect\citeauthoryear{Butcher \& Oemler}{1985}]{bo}
Butcher, H. R., \& Oemler, A. Jr., 1985, ApJS, 57, 665

\bibitem[\protect\citeauthoryear{Carlberg}{1984}]{carlberg}
Carlberg, R. G. 1984, ApJ, 286, 403

\bibitem[\protect\citeauthoryear{Davies, Sadler \& Peletier}{1993}]{dsp}
Davies, R. L., Sadler, E. M., \& Peletier, R. F. 1993, MNRAS, 262, 650

\bibitem[\protect\citeauthoryear{Fioc \& Rocca-Volmerange}{1997}]{fiocrocca}
Fioc, M., \& Rocca-Volmerange, B. 1997, A\&A, 326, 950

\bibitem[\protect\citeauthoryear{Goudfrooij \& de Jong}{1995}]{goud}
Goudfrooij, P. \& de Jong, T. 1995, A\&A, 298, 784

\bibitem[\protect\citeauthoryear{Hawarden et al.}{2001}]{fs}
Hawarden, T. G., Leggett, S. K., Letawsky, M. B., Ballantyne, D. R., \&
Casali, M. M. 2001, MNRAS, 325, 563

\bibitem[\protect\citeauthoryear{Henry \& Worthey}{1999}]{henry}
Henry, R. B. C., \& Worthey, G. 1999, PASP, 111, 919

\bibitem[\protect\citeauthoryear{Kawata}{1999}]{kawata}
Kawata, D. 1999, PASJ, 51, 931

\bibitem[\protect\citeauthoryear{Kawata \& Gibson}{2003}]{kawatagibson}
Kawata, D., \& Gibson, B. K. 2003, MNRAS, 340, 908

\bibitem[\protect\citeauthoryear{Kobayashi}{2004}]{kobayashi}
Kobayashi, C. 2004, MNRAS, 347, 740

\bibitem[\protect\citeauthoryear{Kobayashi \& Arimoto}{1999}]{kobaari}
Kobayashi, C., \& Arimoto, N. 1999, ApJ, 527, 573

\bibitem[\protect\citeauthoryear{Kodama \& Arimoto}{1997}]{ka97}
Kodama, T., \& Arimoto, N. 1997, A\&A, 320, 41

\bibitem[\protect\citeauthoryear{Kodama et al.}{1998}]{kodama98}
Kodama, T., Arimoto, N., Barger, A. J., \& Arag\'{o}n-Salamanca,
A. 1998, A\&A, 334, 99 

\bibitem[\protect\citeauthoryear{Kroupa, Tout \& Gilmore}{1993}]{kroupa}
Kroupa, P., Tout, C. A., \& Gilmore, G. 1993, MNRAS, 262, 545

\bibitem[\protect\citeauthoryear{La Barbera et al.}{2003}]{labarbera}
La Barbera, F., Busarello, G., Massarotti, M., Merluzzi, P., \&
Mercurio, A. 2003, A\&A, 409, 21

\bibitem[\protect\citeauthoryear{Landolt}{1992}]{landolt}
Landolt, A. U. 1992, AJ, 104, 340

\bibitem[\protect\citeauthoryear{Larson}{1974}]{larson}
Larson, R. B. 1974, MNRAS, 169, 229

\bibitem[\protect\citeauthoryear{Lucey et al.}{1997}]{lucey}
Lucey, J. R., Guzm\'{a}n, R., Steel, J., \& Carter, D. 1997, MNRAS,
287, 899

\bibitem[\protect\citeauthoryear{Mehlert et al.}{2003}]{mehlert03}
Mehlert, D., Thomas, D., Saglia, R. P., Bender, R., \& Wegner, G. 2003,
A\&A, 407, 423

\bibitem[\protect\citeauthoryear{Miller \& Scalo}{1979}]{ms}
Miller, G. E., \& Scalo, J. M. 1979, ApJS, 41, 513

\bibitem[\protect\citeauthoryear{Peletier et al.}{1990}]{pela}
Peletier, R. F., Davies, R. L., Illingworth, G. D., Davis, L. E., \&
Cawson ,M. 1990, AJ, 100, 1091 

\bibitem[\protect\citeauthoryear{Peletier, Valentijn \& Jameson}{1990}]{pelb}
Peletier, R. F., Valentijn, E. A., \& Jameson, R. F. 1990, A\&A, 233,
62

\bibitem[\protect\citeauthoryear{Rood \& Sastry}{1972}]{rs}
Rood, H. J., \& Sastry, G. N. 1972, AJ, 77, 451

\bibitem[\protect\citeauthoryear{Saglia et al.}{2000}]{saglia}
Saglia, R. P., Maraston, C., Greggio, L., Bender, R., \& Ziegler,
B. 2000, A\&A, 360, 911 

\bibitem[\protect\citeauthoryear{Simien \& de Vaucouleurs}{1986}]{simien}
Simien, F., \& de Vaucouleurs, G. 1986, ApJ, 302, 564

\bibitem[\protect\citeauthoryear{Stanford, Eisenhardt \& Dickinson}{1998}]{stanford}
Stanford, S. A., Eisenhardt, P. R., \& Dickinson, M. 1998, ApJ, 492,
461 

\bibitem[\protect\citeauthoryear{Tamura et al.}{2000}]{tamura1}
Tamura, N., Kobayashi, C., Arimoto, N., Kodama, T., \& Ohta, K. 2000,
AJ, 119, 2134 

\bibitem[\protect\citeauthoryear{Tamura \& Ohta}{2000}]{tamura2}
Tamura, N., \& Ohta, K. 2000, AJ, 120, 533

\bibitem[\protect\citeauthoryear{Tamura \& Ohta}{2003}]{tamura3}
Tamura, N., \& Ohta, K. 2003, AJ, 126, 596 (Paper I)

\bibitem[\protect\citeauthoryear{Vazdekis et al.}{1996}]{vazdekis}
Vazdekis, A., Casuso, E., Peletier, R. F., \& Beckman, J. E. 1996,
ApJS, 106, 307 (V96)

\bibitem[\protect\citeauthoryear{Wise \& Silva}{1996}]{wise}
Wise, M. W., \& Silva, D. R. 1996, ApJ, 461, 155

\bibitem[\protect\citeauthoryear{Witt, Thronson \& Capuano}{1992}]{witt}
Witt, A. N., Thronson, H. A. Jr., \& Capuano, J. M. Jr. 1992, ApJ, 393,
611

\bibitem[\protect\citeauthoryear{Wu et al.}{2004}]{wu}
Wu, H., Shao, Z., Mo, H. J., Xia, X., \& Deng, Z. 2004, ApJ, submitted
(astro-ph/0404226) 

\end{thebibliography}
\end{document}